# UNUSUAL MICROWAVE RESPONSE AND BULK CONDUCTIVITY OF VERY THIN FeSe$_{0.3}$Te$_{0.7}$ FILMS AS A FUNCTION OF TEMPERATURE

A.A. Barannik[1], N.T. Cherpak[1], Yun Wu[2], Sheng Luo[2], Yusheng He[3], M.S. Kharchenko[1], A. Porch[4]

[1]*O. Ya. Usikov Institute for Radiophysics and Electronics, National Academy of Science, 61085 Kharkiv, Ukraine*

[2]*University of Science and Technology, 100083 Beijing, China Institute of Physics, China*

[3]*ChineseAcademy of Science 100190 Beijing, China*

[4]*CardiffUniversity, Cardiff CF24 3AA, Wales, UK*

Results of X-band microwave surface impedance measurements of FeSe$_{1-x}$Te$_x$ very thin film are reported. The effective surface resistance shows appearance of peak at $T \leq T_c$ when plotted as function of temperature. The authors suggests that the most well-reasoned explanation can be based on the idea of the changing orientation of the microwave magnetic field at a SN phase transition near the surface of a very thin film. The magnetic penetration depth exhibits a power-law behavior of $\delta\lambda_L(T) \propto CT^n$, with an exponent $n \approx 2.4$ at low temperatures, which is noticeably higher than in the published results on FeSe$_{1-x}$Te$_x$ single crystal. However the temperature dependence of the superfluid conductivity remains very different from the behavior described by the BCS theory. Experimental results are fitted very well by a two-gap model with $\Delta_1/kT_c = 0.43$ and $\Delta_2/kT_c = 1.22$, thus supporting s$_\pm$ - wave symmetry. The rapid increase of the quasiparticle scattering time is obtained from the microwave impedance measurements.

## 1. Introduction

The discovery of superconductivity in the Fe-based pnictide compound LaFeAsO$_{1-x}$F ("1111") has stimulated a great scientific interest and intense studies of this class of superconductors [1]. The compounds contain the ferromagnetic element Fe and so unconventional superconducting properties were expected because (in general) superconductivity and ferromagnetism are usually antagonistic. Considerable efforts have been performed in searching for superconductivity in structurally simple Fe-based substances. As a result, the metallic superconductors BaFe$_2$As$_2$ ("122") with Co- and Ni- doping were discovered [2-4].

The discovery of superconductivity in pnictides (e.g. in "1111" and "122") and chalcogenides (e.g. in "11") is of great importance, because it gives additional chance to study nature of superconductivity in these substances and cuprates by means of comparison of their properties. Especially, the discovery of superconductivity in binary As-free Fe-chalcogenide ("11") is of great interest, since it only contains the FeSe-layer, which has an identical structure as FeAs, and the Se deficiency may be the reason of the superconductivity [5]. By introducing Te, the critical temperature in FeSe$_x$Te$_{1-x}$ can be increased. This system is convenient because the doping can be well controlled [6].

For this new family of unconventional superconductors, the pairing symmetry of their energy gap is a key to understanding the mechanism of superconductivity. Extensive experimental and theoretical works have been done to address this important issue for FeAs-based superconductors. At present, increasing evidence points to multi-gap models of superconductivity, possibly with an unconventional pairing mediated by antiferromagnetic fluctuations[7, 8]. Thus new experimental works and theoretical approaches are very important for reliable conclusions.

The measurement of the temperature dependence of the microwave impedance is a powerful tool for studying not only the penetration depth [9] but also the whole complex conductivity of the samples [10]. To date, few works have been published on the experimental study of microwave surface impedance of FeSe-based chalcogenides [11, 12]. The work [11] reports microwave surface impedance of FeSe$_{0.4}$Te$_{0.6}$ single crystals and the power-law behavior of penetration depth $CT^n$ with an exponent $n \approx 2$, which is considered to result from impurity scattering and differs noticeably from $n$ in other Fe-based superconductors, e.g. $n = 2.8$ in Ba(Fe$_{1-x}$Co$_x$)$_2$As$_2$ [13]. The work [12] is the study of very thin epitaxial FeSe$_{0.3}$O$_{0.7}$ film of thickness $d_f$ less than penetration depth $\lambda_L$ in the whole temperature range. In this case some unclear features of the microwave effective surface impedance were observed, depending on the temperature. They are: 1) the appearance of a peak in the effective surface resistance $R_s^{eff}$ at $T \leq T_c$; 2) a considerable difference between the effective film surface resistance $R_s^{eff}$ and reactance $X_s^{eff}$ at $T > T_c$.

This present work is aimed at obtaining bulk (i.e. intrinsic) microwave properties $R_s(T)$ and $X_s(T)$ using our experimental data [12] and thus to obtain temperature dependences of the penetration depth $\lambda_L(T)$, the quasiparticle conductivity $\sigma_1(T)$, the conductivity of the superfluid component $\sigma_2(T)$ and the quasiparticle scattering rate $\tau^{-1}(T)$. These values are then compared with the results obtained for our thin epitaxial film and single crystal [11] of the same compound FeSe$_{1-x}$Te$_x$. Appendix A gives an expression for the effective surface impedance $Z_s^{eff}$ as a function of film thickness $d_f$ in terms of the bulk surface impedance $Z_s$ for three configurations of microwave magnetic field near the surfaces of the sample under study.

## 2. Experimental data and their peculiarities

Epitaxial FeSe$_{1-x}$Te$_x$ ($x = 0.7$) film deposited on a LaAlO$_3$ substrate by a pulsed laser deposition method [14, 15] is found to have $T_c$ $_{onset}$ = 14.8 K and a transition width $\Delta T$ = 1.6K on the levels of resistivity $\rho(T)/\rho(T_c$ $_{onset})$= 0.1 and 0.9 (inset in Fig.1a). The microwave response of the film was measured using an X-band sapphire dielectric resonator. It is a close analogy to [16].

The cavity resonator, which has a quality factor of $Q_0$ = 45000 at room temperature, is specially designed for the microwave measurements of small samples using the TE$_{011}$ mode, with the sapphire cylinder having a small hole along its axis. The sample with film thickness $d_f$ = 100 nm and other, lateral dimensions of 1 mm is put in the center of the hole but isolated from the cylinder, supported by a very thin sapphire rod. The cavity is sealed in a vacuum chamber immersed in liquid $^4$He and the temperature of sapphire rod (hence the sample) can be controlled from 1.6 to 60 K with a stability about ± 1 mK while keeping the cavity at a temperature of 4.2 K. The temperature dependence of resonance frequency and quality factor of resonator (Fig. 1a and 1b) were measured by a vector network analyzer (Agilent N5230C) for both the thin film sample and also the bare substrate.

The effective surface resistance (Fig. 2a) is determined by the expression

$$R_s^{eff} = \frac{Q_s^{-1} - Q_{ws}^{-1}}{A_s}, \qquad (1)$$

where $Q_s$ and $Q_{ws}$ are the $Q$-factors of the resonator with and without the sample under study, respectively. The coefficient of inclusion [17] $A_s = 2.9 \cdot 10^{-4}$ $\Omega^{-1}$ was obtained by modeling using CST 2009. The surface reactance $X_s^{eff}(T)$ can be written as

$$X_s^{eff} = X_s^{eff}(0) + \Delta X_s^{eff}(T), \qquad (2)$$

where $X_s^{eff}(0)$ is the effective reactance at $T = 0$ and

$$\Delta X_s^{eff}(T) = -\frac{2\Delta f(T)}{A_s f_0} \qquad (2a)$$

here $f_0$ is the center frequency of the resonator and $\Delta f(T)$ is the frequency shift relative to the resonator without the sample. $\Delta X_s^{eff}(T)$ is presented in Fig 2b.

As can be seen in Fig. 2, in the temperature dependence of the microwave response $Z_s^{eff} = R_s^{eff} + iX_s^{eff}$ of the film there are two features that were not presented in the study of YBa$_2$Cu$_3$O$_{7-\delta}$ single crystals (see, e.g. [18]) and films (see, e.g. [19, 20]), as well as for FeSe$_{0.4}$Te$_{0.6}$ single crystal [11]. The most noticeable feature is a peak of $R_s^{eff}(T)$ at $T \leq T_c$ near $T_c$. The second feature of the response is manifested in the abnormally large change in the resonant frequency of the resonator and thus in the effective growth of the reactance when approaching $T_c$. The possible nature of these features is discussed below.

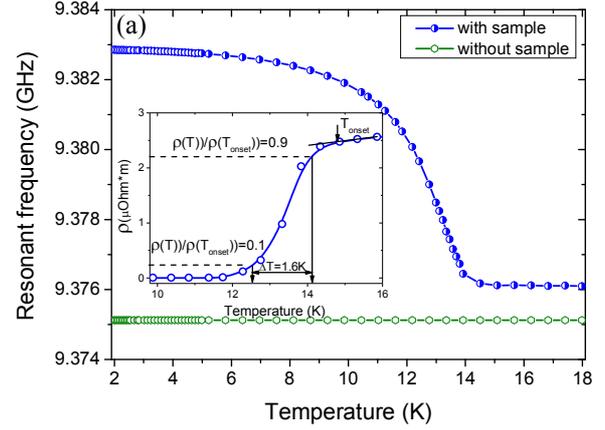

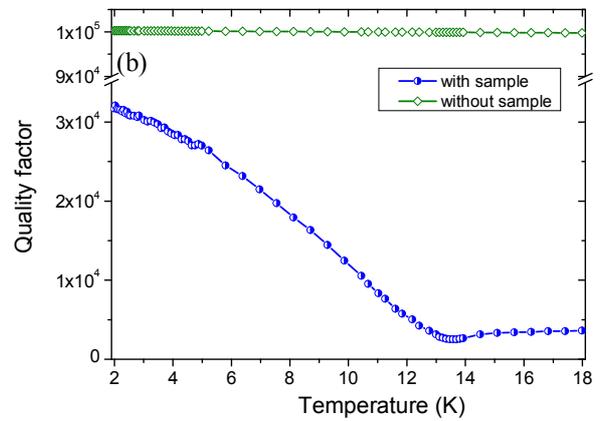

*Fig. 1.* The temperature dependence of resonance frequency (a) and quality factor (b) of the resonator for both the thin film sample and the bare substrate (empty symbols). The inset shows the temperature dependence of the resistivity.

## 3. Finding the bulk surface impedance

The film under study was placed in the resonator so that its plane was perpendicular to the rotational symmetry axis of the resonator, i.e. perpendicular to microwave magnetic field $H_\omega$ in a center of the cavity. It is known that when the film thickness $d_f$ is comparable to the magnetic penetration depth $\lambda_L$ the measured surface impedance is a function of the ratio $d_f/\lambda_L$ [10]. At the same time, relations between $Z_s^{eff}(d_f/\lambda_L)$ and bulk impedance $Z_s$ are known for two cases of configurations of microwave field $H_\omega$ at the surface of the film: 1) field is symmetric with respect to two side surfaces and 2) field has a component on one side of the film [10]. The first case is typical for placing the film in the resonator with HE$_{011}$-mode parallel to the field $H_\omega$, the second case occurs when the film is the conducting endplate of metal or dielectric resonator [17]. In our work the third case is realized, when the magnetic field at the planes of the film is in the opposite directions (see Appendix A). In this case the above mentioned relation has the form

$$Z_s^{eff}(d_f/\lambda_L) = -\frac{i}{2} Z_s \cot\left(k\frac{d_f}{2}\right), \quad (3)$$

where $k = \omega\mu_0/Z_s$, $\omega = 2\pi f$, $\mu_0 = 4\pi \cdot 10^{-7}$ H/m.

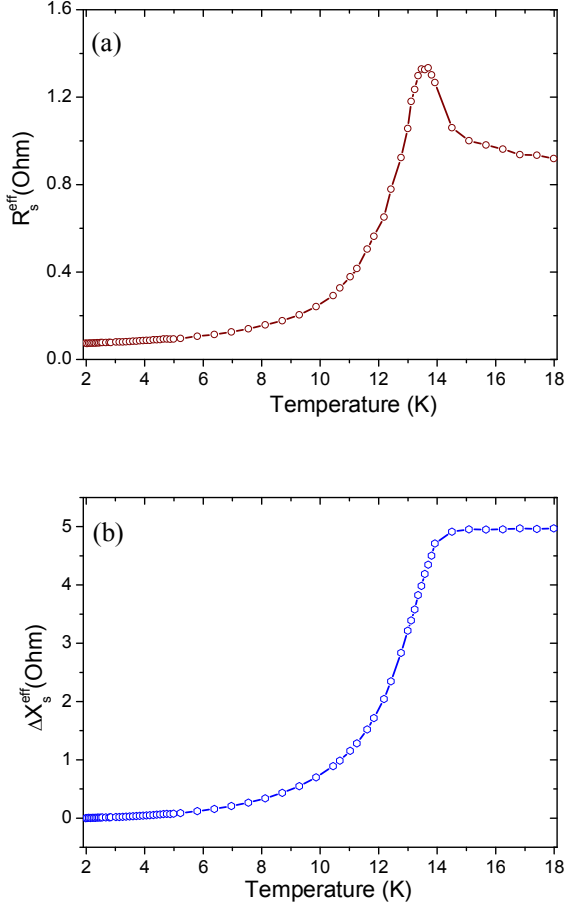

*Fig. 2.* (a) The effective surface resistance $R_s^{eff}$, and (b) change of surface reactance $\Delta X_s^{eff}$ of FeSe$_{1-x}$Te$_x$ film depending on temperature.

Since the penetration depth at $T = 0$, $\lambda_L(0)$, is not determined in our work, for the purpose of finding $Z_s(T)$ we need to use the values of $\lambda_L(0)$, obtained in other works. These values are known, e.g., 470 nm (single crystal FeSe$_{0.4}$Te$_{0.6}$, microwave measurement) [11], 560 nm (single crystal FeTe$_{0.58}$Se$_{0.42}$, TDR measurement) [21] and 534 nm (powder sample of FeTe$_{0.5}$Se$_{0.5}$, μSR) [22]. Obviously, our film is much thinner than $\lambda_L(0)$, and in addition the ratio $d_f/\lambda_L(T)$ further decreases with increasing temperature.

In the case when $R_s \ll X_s$, that is expected in the temperature range from $T = 0$ to $\sim T_c/2$, equation (3) reduces to

$$R_s^{eff} = \frac{1}{2} R_s \left( \coth\left(\frac{d_f}{\lambda_L}\right) + \frac{d_f}{2\lambda_L} \cosec^2\left(\frac{d_f}{2\lambda_L}\right) \right);$$

$$X_s^{eff} = \frac{1}{2} X_s \coth\left(\frac{d_f}{2\lambda_L}\right). \quad (3a)$$

In the limit of very thin films ($d_f/\lambda_L \ll 1$) $R_s^{eff} = 2R_s\lambda_L/d_f$ and $X_s^{eff} = X_s\lambda_L/d_f$.

Expressions (3a) were used to find the bulk (intrinsic) values of $R_s$ and $X_s$ (Fig. 3). There we used the equality $R_s = X_s$ at $T \geq T_c$ and $\lambda_L(0) = 560$ nm [21].

### 4. Discussion of the results

The approach of finding $R_s$ and $X_s$ in the previous section does not explain the nature of the appearance of the peak $R_s^{eff}$ near $T_c$. We can consider several explanations in this respect: 1) the coherence peak, 2) manifestation of the magnetic component in a superconductor ($\mu > 1$), 3) the size effect at $d_f \approx \lambda_L(T)$; 4) the effect of changing the microwave magnetic field configuration near the film surfaces at a S-N transition.

Apparently we should not talk about coherence peak, because it is not observed in the single crystal [11]. The appearance of the magnetic component in a superconductor, when the relative permeability $\mu > 1$, is possible in principle [23]. However, the effect in [23] was observed at $T > T_c$ and the peak width is significantly greater than in the present work.

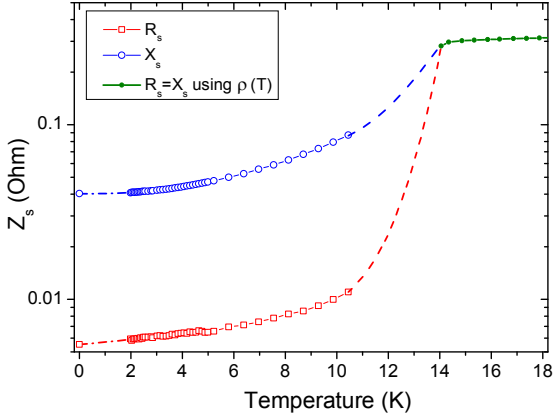

*Fig. 3.* Surface resistance $R_s(T)$ and surface reactance $X_s(T)$

We observe the peak of $R_s^{eff}$ at $T \leq T_c$. The size effect when $d_f$ is comparable with $\lambda_L(T)$ or normal state skin depth is excluded completely because $d_f < \lambda_L(T)$ in the whole temperature interval from $T = 0$ to $T_c$.

It seems that the most well-reasoned explanation can be based on the idea of the changing of the parallel orientation of the microwave magnetic field near the surface of a superconductor at the phase transition from the S-state to an orientation close to perpendicular in the N-state. This occurs when the field direction near the surface of the very thin film coincides at least partially with the direction of TE$_{011}$-mode field near the axis of the resonator (Fig. A1 in Appendix). Here the correlation between $Z_s^{eff}$ and $Z_s$ must change. Evidently, $|Z_s^{eff}| > |Z_s|$ at $T < T_c$ and $|Z_s^{eff}| < |Z_s|$ at $T \geq T_c$ (See Appendix and [10]). We have no mathematical model describing changing $Z_s^{eff}(T)$ and the relationship of $Z_s^{eff}(T)$ and $Z_s(T)$ near $T_c$, therefore we found $R_s(T)$ in the interval of T = 1.6 – 10 K in accordance with (3a) and determined $R_s(T)$ a $T \geq T_c$ using $R_s = \sqrt{\omega\mu_0\rho/2}$, where $\rho$ is the measured resistivity. After that we matched up the obtained values of $R_s$ in a region of $T \leq T_c$. The dependence $X_s(T)$ was found using $X_s(0) = \omega\mu_0\lambda_L(0)$ at $T = 0$, the dependence of $\Delta X_s(T)$ taking into account the expression (3a), and using the equality $R_s = X_s$ at $T \geq T_c$ and the matching described above. The correctness of this approach was validated by the mutual co-ordination of $R_s^{eff}(T)$, $X_s^{eff}(T)$, $R_s(T)$ and $X_s(T)$ within the framework of equation (3). The obtained values of $R_s(T)$ and $X_s(T)$ allow us to find the complex conductivity of the sample, where $\sigma_1$ is the quasiparticle conductivity

$$\sigma_1 = 2\omega\mu_0 \frac{R_s X_s}{|Z_s|^4} \qquad (4)$$

and $\sigma_2$ is the conductivity of the superfluid component

$$\sigma_2 = \omega\mu_0 \frac{X_s^2 - R_s^2}{|Z_s|^4}, \qquad (5)$$

where $|Z_s|^4 = (R_s^2 + X_s^2)^2$.

At low temperature when the condition $\sigma_1 \ll \sigma_2$ is true, we easily obtain from (5) the known expression $\sigma_2 = \omega\mu_0/X_s^2$. Because $\sigma_2 = e^2 n_s/m\omega = 1/\mu_0\omega\lambda_L^2$, it is easy to obtain the well known relation

$$X_s = \mu_0\omega\lambda_L, \qquad (6)$$

which is often used for obtaining information about the structure of the energy gap in superconductors (see e.g. [9]). In general $\lambda_L(T)$ is found from equation (5) as

$$\lambda_L(T) = \frac{1}{\sqrt{\mu_0\omega\sigma_2(T)}}. \qquad (7)$$

The temperature dependence $\lambda_L(T)$ is shown in Fig. 4, where the inset presents the low-temperature part of the data. The absolute value of $\lambda_L(0) = 560$ nm is taken from [21], consistent with other published data [11, 21]. The penetration depth is found to obey a power-low behavior, i.e. $\delta\lambda_L(T) = \lambda_L(T) - \lambda_L(0) \propto CT^n$ with the exponent $n \approx 2.4$ for temperatures as high as 7 K $\geq T_c/2$. The obtained value of $n$ is higher noticeably than $n \approx 2$ in [11, 21] and lower than value of $n$ = 2.8, for example, in crystal Ba(Fe$_{1-x}$Co$_x$)$_2$As$_2$ [24, 13] obtained in the radiofrequency and microwave ranges. Generally speaking, a power-law temperature behavior can be explained by some quantity of low-energy quasiparticles, however it depends also on the presence of magnetic and non-magnetic impurities [11, 25]. Particularly, in a superconductor with s$_\pm$-wave symmetry with non-magnetic impurities the behavior $\lambda_L(T)$ at low temperature has a form of $T^2$.

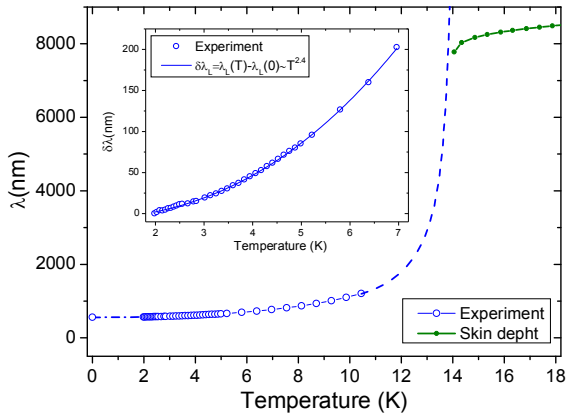

*Fig. 4.* Temperature dependence $\lambda_L(T)$. The inset presents the low temperature part of the dependence $\delta\lambda_L(T) = \lambda_L(T) - \lambda_L(0)$, and the solid line corresponds to power-law behavior $\delta\lambda_L(T) = CT^n$ with $n$ = 2.4.

Fig. 5 displays $\sigma_1(T)$, both with and without subtracting $R_s(0) = R_{res}$ from the bulk data of $R_s$ before calculating $\sigma_1$. Subtracting $R_{res}$ removes the influence of surface defects and so yields the quasiparticle behavior [11]. We should note that an error in the estimate of $R_{res}$ changes noticeably $\sigma_1(T)$ at low temperatures but relatively much less so in the higher temperature part of the S-state. It can be seen in our work with thin films and in [11] with single crystals of very close composition that a considerable enhancement of $\sigma_1(T)$ is observed below $T_c$. Such an enhancement was also observed in cuprate high-$T_c$ superconductors [26, 27, 20] and in Fe-based pnictides [28, 29] and is much broader than the coherence peak. It is explained by suppression of quasiparticle scattering below $T_c$ when quasiparticle density decreases, giving the appearance of the broad peak in $\sigma_1(T)$ below $T_c$.

In this situation it is important to find the quasiparticle scattering rate $\tau^{-1}$ in the system under study. On the assumption that all charge carriers condense at $T$ = 0 and $\omega\tau \ll 1$ the following relation is valid [30]

$$\tau^{-1} = \frac{1 - \lambda_L^2(0)/\lambda_L^2 T}{\mu_0 \sigma_1(T)\lambda_L^2(0)}, \qquad (8)$$

where $\lambda_L(T)$ can be found from (7). To this end we need to find a conductivity $\sigma_2(T)$, which in turn is determined by the equation (5).

Figure 6 shows the temperature dependence of $\sigma_2$ for two values of residual surface resistance. As one can see, the two curves are very close. The obtained data allows us to find $[\lambda_L(0)/\lambda_L(T)]^2$. Figure 7 displays the comparison of experimental results with theoretical models in the low-temperature part of $\lambda_L(T)$ and indicates a very good fit of experiment data with the two-gap model when $\Delta_1$ = 0.43 $kT_c$ (weight coefficient 0.84) and $\Delta_2$ = 1.22 $kT_c$.

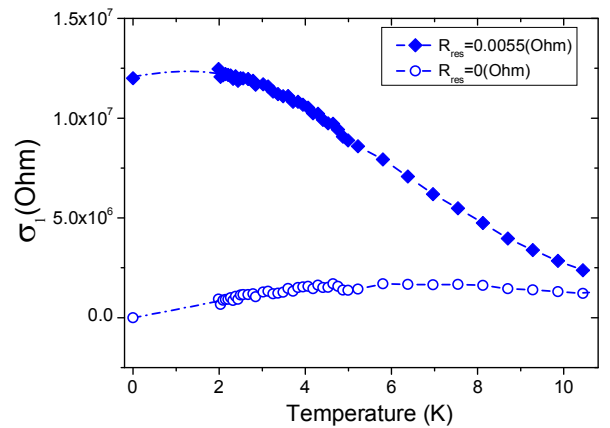

*Fig. 5.* Quasiparticle conductivity $\sigma_1(T)$ obtained with (empty symbols) and without (filled symbols) subtracting residual resistance $R_{res}$

These results noticeably differ from ones in [11] obtained by the microwave technique ($\Delta_1$

$=\Delta_2 = 0.85kT_c$) and [31] obtained by TDR technique ($\Delta_1 = 1.93\ kT_c$ and $\Delta_2 = 0.9\ kT_c$), although support $s_\pm$-wave symmetry of the paired electrons. Obviously, the source of discrepancy can be established with a further study of the compounds.

The obtained data on $\sigma_2(T)$ and $\lambda_L(T)$ allow us to find $\tau^{-1}$ depending on $T$. Fig. 8 shows apparently a common feature in the behavior of all known unconventional superconductors, which consists of a sharp decrease in the quasiparticle scattering rate at low temperatures. If we take $R_{res}<5m\Omega$, the rate $\tau^{-1}(T)$ starts to increase with decreasing temperature, which seems to us unphysical.

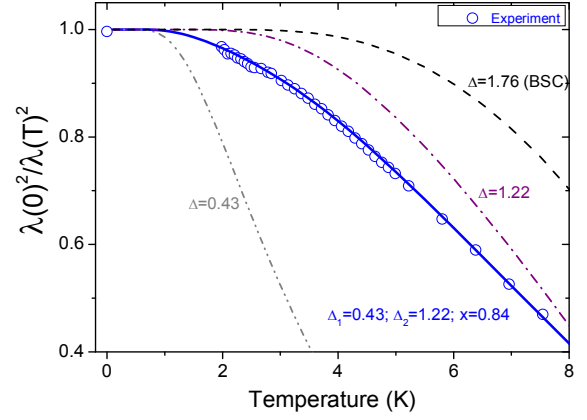

Fig. 7. Ratio $[\lambda_L(0)/\lambda_L(T)]^2$ depending on temperature. The solid line corresponds to the two-gap model ($\Delta_1 = 0.43kT_c$; $\Delta_2 = 1.22\ kT_c$, weight coefficient $x$ is 0.84 for $\Delta_1$), the dashed line corresponds to BCS theory, $\Delta_1 = 0.43kT_c$ (dash-dot-dotted) and $\Delta_2 = 1.22kT_c$ (dash-dotted).

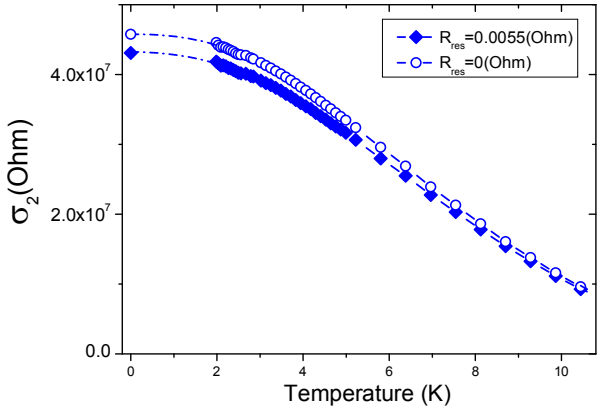

Fig. 6. Conductivity $\sigma_2$ of the superfluid component depending on temperature, taking into account residual surface resistance ($R_{res}$=0.0055 Ohm, lower curve) and without it ($R_{res}$=0, higher curve).

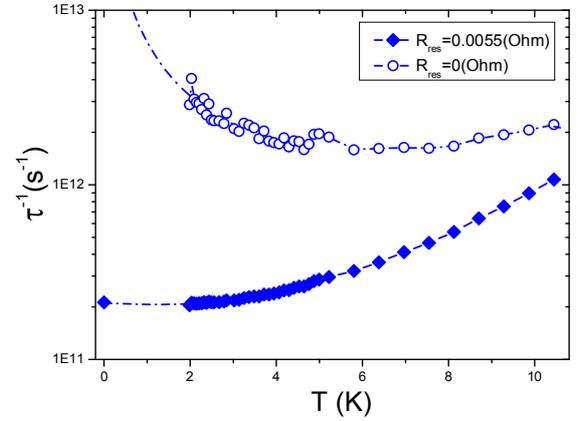

Fig. 8. Quasiparticle scattering $\tau^{-1}(T)$ taking into account $R_{res}$ (filled symbols) and at $R_{res} = 0$ (empty symbols).

The energy gaps found in this work differ markedly from the values shown in other studies (see, eg, [11, 21, 22, 31]). Currently we can not give any convincing explanation for these discrepancies. Obviously, it is necessary to conduct microwave measurements for the film and single crystal FeSe$_{1-x}$Te$_x$ of the same composition using the same resonator(s), and measuring for two alternative orientations of the film in the resonator. It is highly desirable to conduct these measurements not only in the X-band, but at a higher frequency, for example, in the K-band [29]. It is important to clarify the nature of the unusual response at $T \leq T_c$, as well as to establish the consensus values of the energy gaps.

## 5. Conclusion

In conclusion, the microwave surface impedance $Z_s$ of epitaxial $FeSe_{1-x}Te_x$ ($x = 0.7$) film of 100 nm thickness deposited on the $LaAlO_3$ substrate has been measured by an X-band sapphire cavity operating in the $TE_{011}$ mode. The effective surface resistance depending on temperature shows the appearance of a peak at $T \leq T_c$. It can be suggested that the most well-reasoned explanation can be based on the idea of a changing orientation of microwave magnetic field near the surface of a very thin film at a S-N phase transition, when the film thickness is less than $\lambda_L(0)$. The penetration depth shows a power-law behavior $\delta\lambda_L(T) \propto CT^n$, with an exponent $n \approx 2.4$ in the low temperature interval, which is noticeably higher than in the published results on $FeSe_{1-x}Te_x$ single crystal. However the temperature dependence of the superfluid conductivity remains very different from behavior described by the BCS theory. Experimental results indicate very good fit of the theoretical two-gap model with $\Delta_1/kT_c = 0.43$ and $\Delta_2/kT_c = 1.22$, supporting $s_\pm$ - wave symmetry. A rapid increase of the quasiparticle scattering time is obtained from the microwave impedance measurements.

## Acknowledgments

Work is supported partially by IRE NAS of Ukraine under State Project No. 0106U011978 and by the State Agency on Science, Innovations and Informatization of Ukraine under Project No. 01113U004311. Work was performed also within the framework of Agreement of collaboration between IRE NASU and IoP CAS.

## Appendix

A1. The effects of finite sample thickness on the measurements of surface impedance

The standard definition of surface impedance assumes that the superconductor has a thickness much larger than $\lambda$ (or, equivalently, the skin depth $\delta$ when in the normal state). This is evidently not valid for thin superconducting films, where the thickness $d_f$ is typically of the same order of magnitude as $\lambda$, particularly at temperature close to $T_c$. In this case the effective surface impedance $Z_s^{eff}$ measured for samples of finite thickness differs from the intrinsic surface impedance $Z_s$ (i.e. the surface impedance of a infinitely thick sample), and becomes a function of $d_f/\lambda$; the exact dependence on thickness depends on the spatial symmetry of the applied microwave field and the three configurations shown in Fig. A1 are considered here.

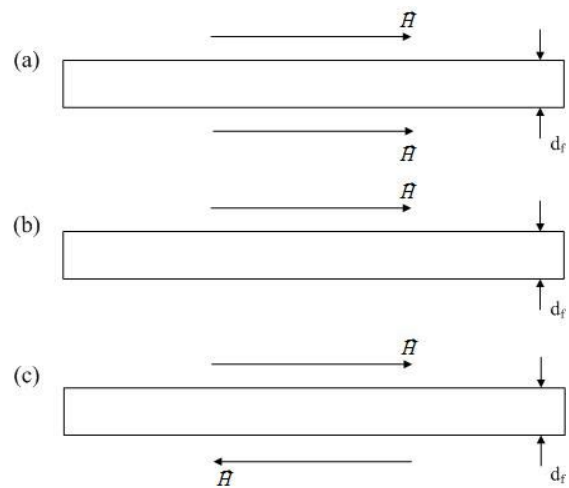

*Fig. A1.* Three possible configurations of microwave field orientation relative to a thin superconducting sample that exhibits

significantly different effective surface impedance in the very thin sample limit; $d_f$ is comparable to $\lambda_L$

Cases (a) and (b) of Fig. A1 are presented in [10] without mathematical details. Here mathematical derivation of the effective surface impedance as a function of $d_f/\lambda$ is given for all three cases.

### A2. Effective surface impedance of thin film

For the case shown in Fig. A2 with $H$ applied parallel to both plane surfaces of film of thickness $d_f$ (see Fig.1(a)), from $\vec{\nabla} \times \vec{H} \approx \vec{J}$ we obtain $\frac{\partial H}{\partial z} \approx \sigma E_y$ if $\vec{H} = (H_x, 0, 0)$ and $\vec{E} = (0, E_y, 0)$. By analogy from $\vec{\nabla} \times \vec{E} = -\frac{\partial \vec{B}}{\partial t}$ we obtain $\frac{\partial E_y}{\partial z} = j\omega\mu_0 H_x$ i.e. $\frac{\partial H_x}{\partial z^2} = j\omega\mu_0 \sigma H_x \equiv -k^2 H_x$.

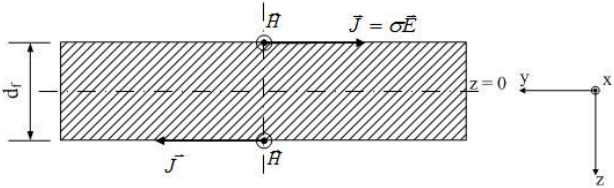

*Fig. A2.* Orientations of microwave field and density current relative to a thin superconducting sample for the case in Fig. A1a).

For the above geometry, we take the symmetric solution for $H_x$, i.e. $H_x(z) = H_x(-z)$ in the form of $H_x = A\cos(kz)$, therefore $E_y = \frac{1}{\sigma}\frac{\partial H_x}{\partial z} = -\frac{Ak}{\sigma}\sin(kz)$. At the film surface where $z = + d_f/2$

$$Z_s^{eff} = \frac{E_y}{H_x}\bigg|_{z=+d_f/2} = -\frac{k}{\sigma}\tan\left(\frac{kd_f}{2}\right)$$

(we get the same value for $Z_s^{eff}$ when $z = -d_f/2$). But $\frac{k}{\sigma} = -\frac{i\omega\mu_0}{k} = -iZ_s$, therefore

$$Z_{eff} = iZ_s \tan\left(\frac{kd_f}{2}\right) \quad (A1)$$

where $Z_s = R_s + iX_s$ is bulk surface impedance. When, $R_s \ll X_s$, $X_s \approx \omega\mu_0\lambda_L$

$$k = \frac{\omega\mu_0}{Z_s} \approx \frac{\omega\mu_0 R_s}{X_s^2} - i\frac{\omega\mu_0}{X_s}$$

$$= \frac{R_s}{\omega\mu_0\lambda_L^2} - i\frac{1}{\lambda_L} = k_1 + ik_2$$

i.e., $k_1 = R_s/\omega\mu_0\lambda_L^2$, $k_2 = -1/\lambda_L$.

We now write $\tan\left(\frac{kd_f}{2}\right) = \tan(\Theta_1 + i\Theta_2)$, where $\Theta_1 = \frac{R_s d_f}{2\omega\mu_0\lambda_L^2}$, $\Theta_2 = -\frac{d_f}{2\lambda_L}$, which is transformed into

$$\tan\left(\frac{kd_f}{2}\right) \approx \frac{\Theta_1 + i\tanh(\Theta_2)}{1 - i\Theta_1\tanh(\Theta_2)}$$

$$\approx \Theta_1 \operatorname{sech}^2\Theta_2 + i\tanh\Theta_2 \quad (A2).$$

Thus by application of (A2) to (A1), $Z_s^{eff}$ can be expressed in terms of $Z_s$ and $d_f/\lambda_L$:

$$Z_s^{eff} = R_s^{eff} + iX_s^{eff} \quad (A3)$$

where

$$R_s^{eff} = \operatorname{Re}(Z_s^{eff}) = R_s\left[\tanh\left(\frac{d_f}{2\lambda_L}\right) - \frac{d_f}{2\lambda_L}\operatorname{sech}^2\left(\frac{d_f}{2\lambda_L}\right)\right],$$

$$X_s^{eff} = \operatorname{Im}(Z_s^{eff}) = X_s \tanh\left(\frac{d_f}{2\lambda_L}\right).$$

For a very thin film $d_f \ll \lambda_L$ write $x = d_f/\lambda_L$ and $\tanh\left(\frac{x}{2}\right) \approx \frac{x}{2}\left(1 - \frac{x^2}{12}\right)$ and $\frac{x}{2}\sec h^2\left(\frac{x}{2}\right) \approx \frac{x}{2}\left(1-\frac{1}{4}x^2\right)$, which then yields the effective surface impedance

$$R_s^{eff} \approx \frac{R_s d_f^3}{12\lambda_L^3} \text{ and } X_s^{eff} \approx X_s \frac{d_f}{2\lambda_L}. \quad (A4)$$

This means that the cylindrical resonator technique for very thin sample lacks sensitivity particularly for measurements of surface resistance when the microwave field is in the same direction on opposite faces of the crystal.

An alternative configuration is case of Fig. A1(b), which occurs when a sample replaces the end-wall of an empty (i.e. air or gas-filled) cylindrical resonator or when a dielectric resonator is used. Both of these resonators are ideal for measuring thin films. The effective surface impedance when $\sigma_1 \ll \sigma_2$ is [31]

$$Z_s^{eff} = R_s\left(\coth x + x\cosech^2 x\right) + iX_s \coth x. \quad (A5)$$

Again defining $x = d_f/\lambda$, we obtain in the film sample limit ($x \ll 1$)

$$R_s^{eff} \approx 2R_s/x; X_s^{eff} = X_s/x. \quad (A6)$$

Therefore, for measuring thin samples it is best to have the microwave field configured as in case b) of Fig. A1. Beyond $x = d_f/\lambda > 3$ the effective and intrinsic surface impedance differ little from each other.

A configuration (c) opposite to case (a) in Fig. A1 occurs, when the microwave field is in opposite directions on opposite faces of the film. By analogy with Fig. A1 for the case (a) we can provide the relative position of H and E vectors, as shown in Fig A3.

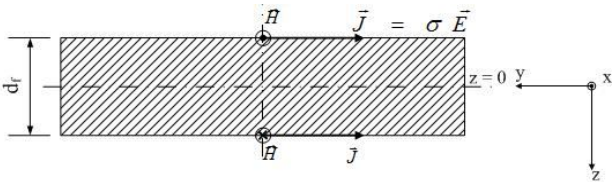

*Fig. A3.* Orientations of microwave field and density current relative to a thin superconducting sample for a case of Fig.A1c).

Here $H_x(z) = -H_x(-z)$. For the given geometry we must take the asymmetric solution for $H_x$ i.e. $H_x = B\sin kz$, therefore $E_y(z) = \frac{Bk}{\sigma}\cos kz$. At the film surface where $z = +d_f/2$

$$Z_s^{`eff} = \frac{k}{\sigma}\cot\left(\frac{kd_f}{2}\right),$$

therefore

$$Z_s^{eff} = -iZ_s \cot\left(\frac{kd_f}{2}\right) \quad (A7)$$

because $\frac{k}{\sigma} = -iZ_s$ (see case a) in Fig. A1). By analogy to case a) we obtain when $R_s \ll X_s$

$$\cot\left(\frac{kd_f}{2}\right) = \Theta_1 \frac{1}{sh^2\frac{d_f}{2\lambda_L}} + i\frac{1}{\tanh\frac{d_f}{2\lambda_L}}.$$

The configuration shown in Fig. A3 is a parallel connection of two conductors, so we can write

$$Z_s^{eff} = \frac{1}{2}Z_s^{`eff} = R_s^{eff} + iX_s^{eff} \quad (A8)$$

where

$$R_s^{eff} = \frac{1}{2}R_s\left(\coth\frac{d_f}{2\lambda_L} + \frac{d_f}{2\lambda_L}\cosech^2\frac{d_f}{2\lambda_L}\right),$$

$$X_s^{eff} = \frac{1}{2}X_s \coth\frac{d_f}{2\lambda}.$$

For thin sample limit ($d_f \ll \lambda_L$) $R_s^{eff} \approx 2R_s \frac{\lambda_L}{d_f}$ and $X_s^{eff} \approx X_s \frac{\lambda_L}{d_f}$, which coincide with the case of Fig. A1b).

Hereinafter the question arises, under what conditions is the configuration in Fig. A1(c) realized? For this purpose we consider the cavity in the $TE_{011}$ mode and with a very thin superconductor sample located perpendicular to the microwave field (Fig. A4).

The microwave magnetic field near the sample in superconducting state has a configuration as shown in Fig A5a).

Fig. A5 shows that the directions of the magnetic field lines are opposite at the top and bottom surfaces of the sample. In this case, the currents have the same orientation in both planes (see Fig A5 a, b, c).

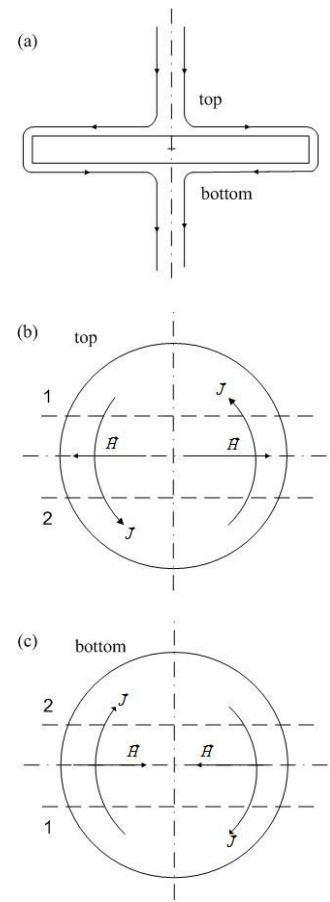

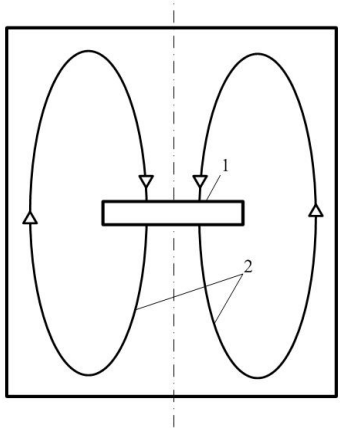

*Fig. A4.* Microwave field configuration in the resonant cavity with $TE_{011}$ mode.

A part of the sample between the planes of cross-sections 1 and 2 with the directions of the magnetic field and currents looks as shown in Fig. A3.

*Fig. A5.* Directions of the magnetic field lines $H$ near a superconducting film placed in the center of a resonator cavity in its $TE_{011}$ –mode (a), (b),(c) and current density $J$ in the top (b) and bottom (c) surface layers of the film.